\documentclass[twocolumn,showpacs,superscriptaddress,prd,showkeys,nofootinbib]{revtex4}
\usepackage{amsmath}
\usepackage{amsfonts}
\usepackage{amssymb}
\usepackage[dvips]{graphicx}
\usepackage{hyperref}
\usepackage{units}
\usepackage{color}
\usepackage{slashed}

\def\b{\begin{equation}}
\def\e{\end{equation}}

\def\la{\langle}
\def\ra{\rangle}
\def\ba{\begin{align}}
\def\ea{\end{align}}
\begin{document}

\title{Can a charged decaying particle serve as an ideal clock in the presence of the magnetic field?}

\author{Roberto Pierini}
\email{rpierini@fuw.edu.pl}
\author{Krzysztof Turzy\'nski}
\email{Krzysztof.Turzynski@fuw.edu.pl} 
\author{Andrzej Dragan}
\email{dragan@fuw.edu.pl}      
\affiliation{Institute of Theoretical Physics, University of Warsaw, 05-093 Warsaw, Poland}


\begin{abstract}
We investigate a model of a supposedly ideal clock based on the decay rate of a charged particle in circular motion in a constant magnetic field.
We show that the time measured by an ideal clock depends on the acceleration. However, the effect becomes visible at an order of magnitude of $10^{28}$ g, therefore confirming the validity of the ideal clock hyphotesis for realistic accelerations.  
\end{abstract}
\pacs{03.30+p, 03.70.+k}
\keywords{ideal clock, proper time, particle decay}
\maketitle
\date{today}

\section{Introduction}
Proper time is defined as the time measured by an ideal clock along its own path. In general, the proper time measured by such a clock moving with instantaneous velocity $v(t)$ along any path is given by the equation
\b\label{propertime}
\tau=\int\sqrt{1-\frac{v^2(t)}{c^2}}\,dt  \,,
\e 
where $dt$ is the ticking rate of ideal clocks at rest. 
For constant speed, $(\ref{propertime})$ reduces to the more familiar time dilation expression, $\tau=t/\gamma$.
So called good clocks, or ideal clocks, are devices that measure time according to the relation $(\ref{propertime})$ along any path, independently on the acceleration they are undergoing. 

Whether a clock is good or not depends on the circumstances of its realization and any conceivable device has to be reconducted to some fundamental physical process. An often invoked, textbook model of an ideal clock consists in an ensemble of decaying unstable particles. Much effort has been invested in verifying this model.  
In 1977, the lifetime of a relativistic muon undergoing a circular motion was measured \cite{Bay}. No evidence of deviation from $(\ref{propertime})$ for high transverse acceleration has been found: "the predictions of special relativity obtain even under accelerations as large as  $10^{18}\,g$ and down to distances less than  $10^{-15}$ cm"  \cite{Bay}. This experiment is cited in standard textbooks, see for example \cite{Har}, as the evidence for the ideal clock hypothesis. Even if clocks can be considered ideal under the regime in which particle accelerators normally operate, more generally, it has been proven  by several authors that this is not the case \cite{Mue,VM,FVM,SY}. They have studied the decay law of different particles, in rectilinear and circular motion, as a function of the proper acceleration $a$ and they have shown that the rate depends on the particle trajectory. In the work \cite{LDL}, an analogous result was interpreted as the evidence that a timing rate is not independent on the acceleration and that ideal clocks are only a convenient fiction. All of those studies are similar in one respect: the agent forcing the particle to accelerate was not specified.
Here, we extend those results to a more realistic physical model, where a charged particle is forced by a constant magnetic field to move along a circular trajectory.\footnote{After completion of this work the following paper was brought to our attention, in which the author non-quantitatively investigates the considered question: \cite{Eis}.} Some of the most important experiments measuring time dilation were carried out through muon's decay, see for example \cite{FS} regarding detection of cosmic muons, besides the already cited \cite{Bay}. Therefore, we would like to investigate the validity of special relativistic time dilation at large accelerations by studying the lifetime of a muon in circular motion through a constant magnetic field. 
Since the calculations turn out to be cumbersome for the three-body fermion's decays relevant for real muons, we simplify the model restricting ourself to two-body decays involving only scalar particles. The paper is organized as follows: in Sec. $II$ we describe the physical model simply listing standard results regarding the motion of a scalar particle in a constant magnetic field. In Sec. $III$ we explicitly calculate the decay rate and show that it deviates from the inertial decay rate at high accelerations. We discuss our findings in Sec. $IV$. Finally, we conclude with some remarks in Sec. $V$.

We work in natural units, $\hbar=c=1$.


\section{Physical Model}\label{PM}
In this section, we briefly recapitulate some basic facts about the classical and the quantum theory of charged particles moving in a constant magnetic field. Details can be found in standard textbooks, see for example \cite{GB}. 
Consider a particle of electric charge $e=-|e|$  and mass $M$ in motion under the influence of a constant magnetic field $\vec B=(0,0,B)$. A possible choice for the four-vector potential is
\b\label{4-vec}
A^{\mu}=(0,0,x\,B,0) \,.
\e
In this setting, a classical particle of energy 
\b\label{ene-class}
E=\sqrt{M^2+(eBR)^2+k_z^2} \,,
\e
can move on the $xy$ plane along a circle of radius
\begin{equation}\label{rad-clas}
R=\frac{p_{\perp}}{eB}\,, 
\end{equation}
where $ p_{\perp}=\gamma\,M\,v_{\perp}$ is the value of the transverse momentum and $v_{\perp}=\sqrt{v_x^2+v_y^2}$ is the  value of the transverse velocity of the particle.                              

The classical radial energy is defined as 
\b
p_{\perp}=eBR \,.
\e
The value of the centripetal acceleration is proportional to the magnetic field and to the transverse momentum of the particle
\b\label{cen-acc-c}
a_{\perp}=\frac{|e|B}{\gamma^2\,M^2}\,p_{\perp}\,,
\e 
where $\gamma=E/ M$ is the Lorentz factor.

The quantum theory of a charged particle in a constant magnetic field resembles that of the quantum harmonic oscillator, even though the dynamics of relativistic scalar particles is governed by the Klein-Gordon equation
\b\label{KG}
(\hat p_{\mu}\hat p^{\mu}-M^2)\phi=0  \,,
\e
with the so-called kinetic momentum operator given by 
\b
\hat p_{\mu}=i\,\partial_{\mu}-eA_{\mu}  \,,
\e
and $\phi$ is the canonically normalized wave function of the particle. 
With the gauge choice $(\ref{4-vec})$, solutions of the equation $(\ref{KG})$ can be written as  
\b\label{wf}
\phi_{k,n}(\vec x,t)=I_n(\rho)\,e^{-i\omega_n t}\,e^{i(k_{y}y+k_zz)} \,,
\e
where the index $k$ stands for $k_y$ and $k_z$ and $n$ are the so-called Landau levels. The function $I_n(\rho)$ satisfies the differential equation 
\b\label{hoe} 
\frac{\partial^2}{\partial\rho^2}I_n(\rho)+\left(\lambda-\rho^2\right)I_n(\rho)=0 \,,
\e
which is equivalent to the Schr$\ddot{\text{o}}$dinger equation for the harmonic oscillator. Here, we have that  
\begin{align}\label{rho}
\rho=\sqrt{|e|B}\left(x+\frac{k_y}{|e|B}\right) \,,
\end{align}
and
\b
\lambda=\frac{\omega^2-M^2-k_z^2\pm|e|B}{|e|B} \,.
\e
Bounded solutions to $(\ref{hoe})$ exist only when $\lambda=2\,n+1$ and $n$ is an integer. This has an effect on the energy eigenvalues:  in the quantum theory they are discrete and depending on $n$ as
\begin{align}\label{ene-quant}
\omega_n&=\sqrt{M^2+(2n+1)|e|B+k_z^2} \,.
\end{align}
Finally, properly normalized solutions are
\b\label{wtf}
I_n(\rho)=\left(\frac{\sqrt{|e|B}}{\sqrt{\pi}\,2^n\,n!}\right)^{1/2}\,e^{-\rho ^2/2}\,H_n(\rho) \,,
\e
with $H_n(\rho)$ being Hermite polynomials.

Knowing the wave function $(\ref{wf})$, we can compute the average value of the particle position squared along the $\hat x$ axes
\b
\langle x^2 \rangle=\frac{2n+1}{|e|B}+\frac{k_y}{|e|B} \,.
\e
This result can be interpreted as 
\b\label{rad-vs-Llev}
R^2=\frac{2n+1}{|e|B} \quad \text{and} \quad x^2_0=\frac{k_y}{|e|B} 
\e
being the radius and the center of the circular trajectory squared, respectively. Note that this interpretation can also be deduced from the equation $(\ref{rho})$.
\section{Decay Rate}\label{DR}
The decaying muon is often offered as an example of an ideal clock. Here we are actually studying a simplified model of muon decay - with only two daughter particles and involving scalar fields only
\b\label{s-ss}
\Phi_{\mu}\to\Phi_{e}\Phi_{\nu} \,
\e
The free field $\Phi_{\nu}$ is quantized in terms of a complete set of the solutions of the Klein-Gordon equation 
\b
\Phi(x)=\sum_{k}\frac{1}{\sqrt{2\,V\,\omega_{k}}}\left(a_{k}e^{-ik_{\lambda}x^{\lambda}}+a^{\dagger}_{ k}e^{ik_{\lambda}x^{\lambda}}\right)  \,,
\e
with
$$
\omega^{\nu}_k=\sqrt{M_{\nu}^2+{}^{\nu}k^2_x+{}^{\nu}k^2_y+{}^{\nu}k^2_z} \,. 
$$
Here, the sum over $k$ means the sum over $k_x$, $k_y$ and $k_z$. 
The mode decomposition for the two charged particle fields $\Phi_{\mu}$ and $\Phi_{e}$ reads 
\b
\Phi(x)=\sum_{n}\sum_k\sqrt{\frac{1}{2\,L_yL_z\,\omega_{n}}}\left(a_k\phi_{n,k}^++b^{\dagger}_{k}\phi_{n,k}^-\right)  \,,
\e

where $k$ stands for $k_y$ and $k_z$, $n$ are the Landau levels and $\omega_{n}$ is given in $(\ref{ene-quant})$. 
 The quantization is carried out in a box of dimensions $L_x$, $L_y$, $L_z$ and with volume $V=L_xL_yL_z$.
The momentum in the $\hat x$ direction is not a good quantum number since it is not conserved and, therefore, does not label the frequency modes of the charged particles.
The interaction Lagrangian leading to the decay process (\ref{s-ss}) is 
\b\label{interaction}
\mathcal{L}_I=-G\,\Phi_{\nu}\,\Phi^{\dagger}_e\,\Phi_{\mu}  \,,
\e
which, in the first order of perturbation theory, results in the following differential transition probability per unit time
\begin{widetext}
\begin{eqnarray} \label{diff-decay-rate}
d\Gamma&=&\frac{L_yL_z\,dk^e_y\,dk^e_z}{(2\pi)^2}\,\frac{V\,d^3k^{\nu}}{(2\pi)^3}\,\times\frac{\left|\la\,k^e\,k^{\nu}|S_I|k^{\mu}\ra \right|^2}{T}   \nonumber \\
&=&\frac{G^2\,|e|B}{8(2\pi)^2}\,\frac{d^2k_e\,d^3k_{\nu}}{\omega^e_n\,\omega_m^{\mu}\,\omega_{\nu}}\,\delta(k^{\mu}_{y}-k_y^e-k_y^{\nu})\,\delta(k^{\mu}_{z}-k_z^e-k_z^{\nu})\,\delta(\omega_m^{\mu}-\omega_n^e-\omega^{\nu})\,\left|A_{n,m}\right|^2  \,,
\end{eqnarray}
\end{widetext}
where 
\b\label{xint}
A_{n,m}=\int_{-\infty}^{\infty}dx\,e^{-ik^{\nu}_xx}\,I_m(\rho^{\mu})I_n(\rho^{e})   \,.
\e
The integers $m$ and $n$ label the Landau levels of the ingoing and of the outgoing charged particles respectively. Note that the Dirac delta functions appearing in the second row of $(\ref{diff-decay-rate})$ express the conservation of energy and momentum in the $y$ and $z$ directions. 
We should mention also that the following formulas have been used, which hold when the dimensions of the box $V$ and the measuring time $T$ are sent to infinity
\begin{align}
&\lim_{L_y\to\infty}|\delta(k_y)|^2=\lim_{L_y\to\infty}\frac{L_y}{2\pi}\,\delta(k_y) \,, \\
&\lim_{L_z\to\infty}|\delta(k_z)|^2=\lim_{L_z\to\infty}\frac{L_z}{2\pi}\,\delta(k_z)   \,,\\
&\lim_{T\to\infty}|\delta(\omega)|^2=\lim_{T\to\infty}\frac{T}{2\pi}\,\delta(\omega) \,.
\end{align} 


The integration over $x$ in $(\ref{xint})$ can be done in the following way \cite{GR}
\begin{align}\label{xintegration}
&\int_{-\infty}^{\infty}dx \,e^{\pm ik_1x}\,e^{-({}^e\rho^2/2)}e^{-({}^{\mu}\rho^2/2)}H_{n}({}^{e}\rho)H_m({}^{\mu}\rho)\nonumber\\
&=\frac{e^{-c/4|e|B}}{\sqrt{|e|B}}\int_{-\infty}^{\infty}d\rho \,e^{-\rho^2}H_{n}(\rho+a)H_m(\rho+b) \nonumber\\
&=\left\{
\begin{array}{rl}
&\frac{e^{-c/4|e|B}}{\sqrt{|e|B}}\,2^m\sqrt{\pi}\,n!\,b^{(m-n)}\,L^{m-n}_n(-2ab) \,,\qquad n\leq m \,,  \\
&\frac{e^{-c/4|e|B}}{\sqrt{|e|B}}\,2^n\sqrt{\pi}\,m!\,a^{(n-m)}\,L^{n-m}_m(-2ab) \,,\qquad m\leq n \,,
\end{array}\right.
\end{align}
where
$L_i^j(z)$ are associated Laguerre polynomials, 

and 
\begin{align}
a&=-\frac{1}{2\sqrt{|e|B}}(k_y^e-k_y^{\mu}\mp ik_{\nu}) \\
b&=\frac{1}{2\sqrt{|e|B}}(k_y^e-\,k_y^{\mu}\pm ik_\nu)=-a^* \\
c&=(k_y-k_y^{\mu})^2\pm 2ik_{\nu}(k_y^e+k_y^{\mu})+k_{\nu}^2  \,.
\end{align}
Therefore, using the formula $(\ref{xintegration})$ togheter with the wave functions $(\ref{wtf})$, the modulus square of (\ref{xint}) can be written in a compact form as
\begin{widetext}
\begin{align}\label{A-funct-mod}
\left|A^{\pm}_{n,m}\right|^2=\frac{e^{-\frac{(k_y^e-k_y^{\mu})^2+k_{\nu}^2}{2|e|B}}}{|e|B}\,\left(\frac{m!}{n!}\right)^{\text{sign}(n-m)}\,\left(\frac{(k_y^e-k_y^{\mu})^2+k_{\nu}^2}{2|e|B}\right)^{|n-m|}\,\left[L^{|n-m|}_{\text{min}(n,m)}\left(\frac{(k_y^e-k_y^{\mu})^2+k_{\nu}^2}{2|e|B}\right)\right]^2   \,,
\end{align}
\end{widetext}
where $\text{min}(n,m)$ gives the smallest between the two indexes and $\text{sign}(n-m)$ is equal to one when $n>m$ and equal to minus one when $n<m$. 
Inserting the expression (\ref{A-funct-mod}) into the equation (\ref{diff-decay-rate}) and carrying out the integration over $d^3k^{\nu}$ and $dk^e_y$, gives the following expression for the decay rate
\begin{widetext}
\begin{eqnarray}\label{decayrate}
\Gamma&=&\frac{G^2}{16\pi}\sum_{n}\,\left(\frac{m!}{n!}\right)^{\text{sign}(n-m)}\frac{1}{\omega_m^{\mu}}\nonumber\\
&\times&\int{dk_z^e\,\frac{e^{-\frac{(\omega_m^{\mu}-\omega_n^e)^2-{}^ek_z^2}{2|e|B}}}{\omega^e_n}\left(\frac{(\omega_m^{\mu}-\omega_n^e)^2-{}^ek_z^2}{2|e|B}\right)^{|n-m|}\left[L^{|n-m|}_{\text{min}(n,m)}\left(\frac{(\omega_m^{\mu}-\omega_n^e)^2-{}^ek_z^2}{2|e|B}\right)\right]^2} \,.
\end{eqnarray}
\end{widetext}
Now, the fact that the number of final states accessible to the emitted charged particle are limited implies 
\b\label{kz-lim}
|k_z^e|\leq\frac{\omega_{\mu}^2-M_e^2-(2n+1)|e|B}{2\,\omega_{\mu}} \,,
\e 
and
\b\label{n-lim}
n\leq\frac{\omega_{\mu}^2-|e|B-M_e^2}{2|e|B}\,.
\e
The expression $(\ref{decayrate})$ together with $(\ref{kz-lim})$ and $(\ref{n-lim})$ is the decay rate of a scalar particle decaying into two scalars in a constant magnetic field. We evaluate it numerically.

In the absence of magnetic field the interaction Hamiltonian (\ref{interaction}) leads to the following decay rate
\b\label{free-decay-rate}
\Gamma'_0=\frac{G^2}{16\,\pi\,M_{\mu}}\left(1-\frac{M_e^2}{M_{\mu}^2}\right)  \,,
\e
where $'$ denotes the particle rest frame. 
The lifetime $\tau_0$ of a moving particle is equal to the lifetime at rest $\tau_0'$ multiplied by the $\gamma$ factor
\b\label{lifetime}
\tau_0=\gamma\,\tau'_0 \,,
\e   
where $\gamma=\omega/M$. The inverse of $(\ref{lifetime})$ corresponds to the decay rate
\b\label{free-decay-rate}
\Gamma_0=\frac{1}{\gamma}\,\Gamma'_0 \,.
\e
The decay law $(\ref{decayrate})$ of a particle moving with energy $\omega_l$ under the influence of a magnetic field $\vec B$, has to converge to the value $(\ref{free-decay-rate})$ when $B$ goes to zero. In the following section we show that this is indeed the case and that deviation from the value $(\ref{free-decay-rate})$ becomes significant at very high accelerations.   


\section{Results}\label{R}    
\begin{figure}[ptb]
\,\,\,\,\qquad\qquad\includegraphics[width=4in]{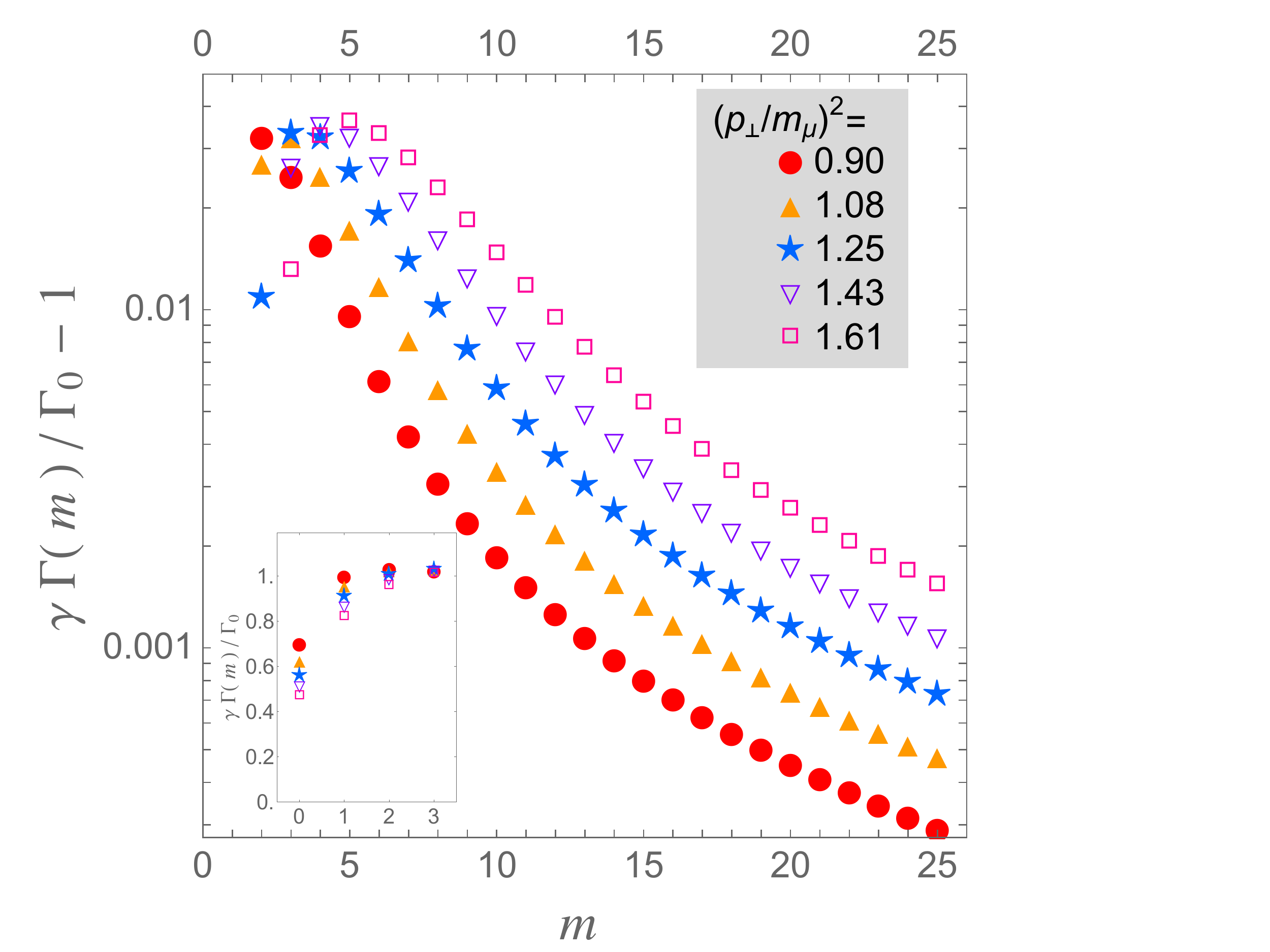}
\caption{Decay rate $\gamma\,\Gamma(m)/\Gamma'_0$ vs Landau levels $m$ for different radial energy values of the initial particle. 
Here $M_{\mu}=105.7$ MeV, $M_e=M_{\nu}=0$.}  
\label{fig1}
\end{figure}

\begin{figure}[ptb]
\includegraphics[width=4.15in]{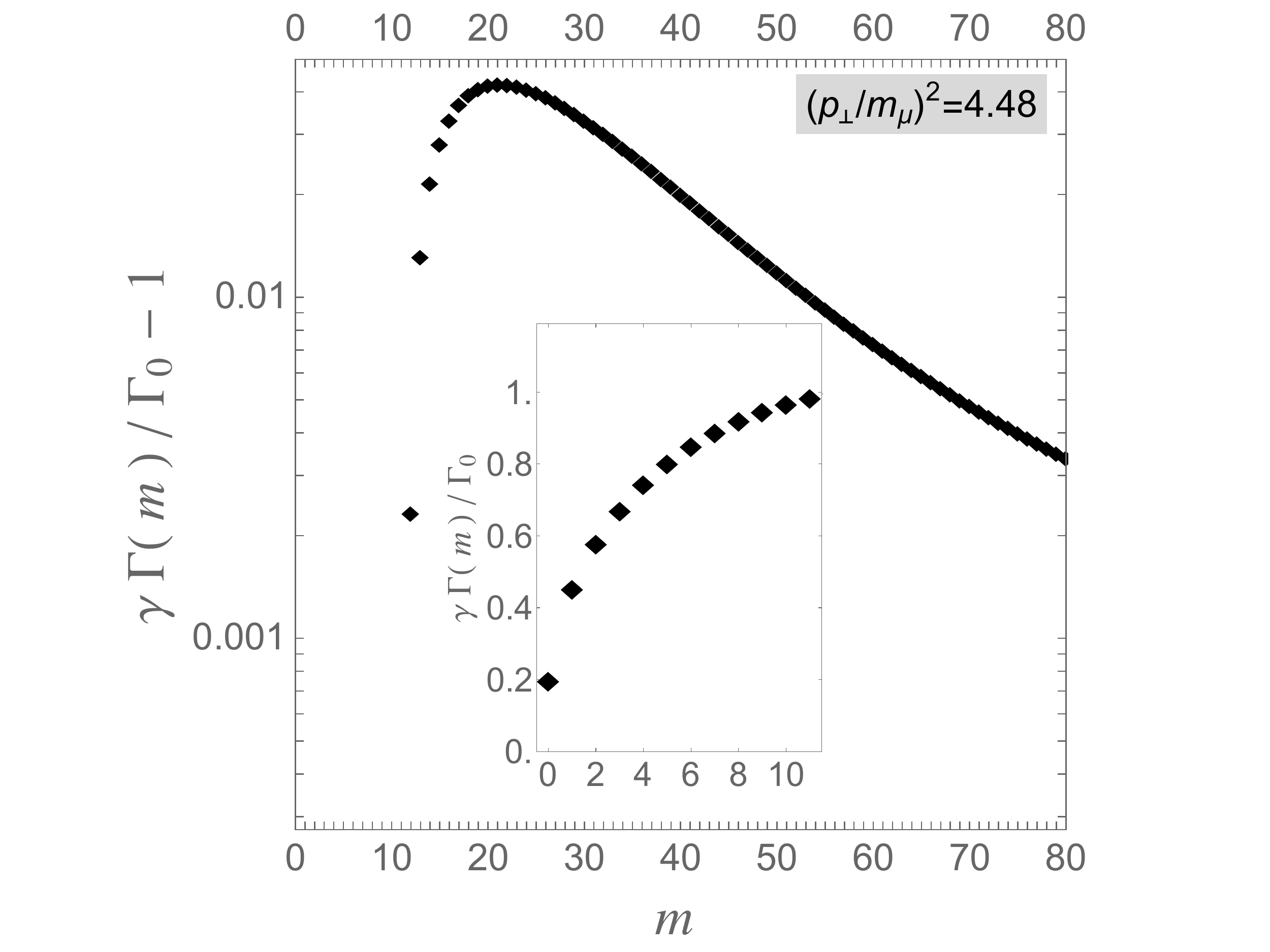}
\caption{Decay rate $\gamma\,\Gamma(m)/\Gamma'_0$ vs Landau levels $m$ for initial radial energy squared $p_{\perp}^2=5\times 10^4 \,\text{MeV}^2$. 
Here $M_{\mu}=105.7$ MeV, $M_e=M_{\nu}=0$.}
\label{fig2}
\end{figure}

\begin{figure}[ptb]
\includegraphics[width=3in]{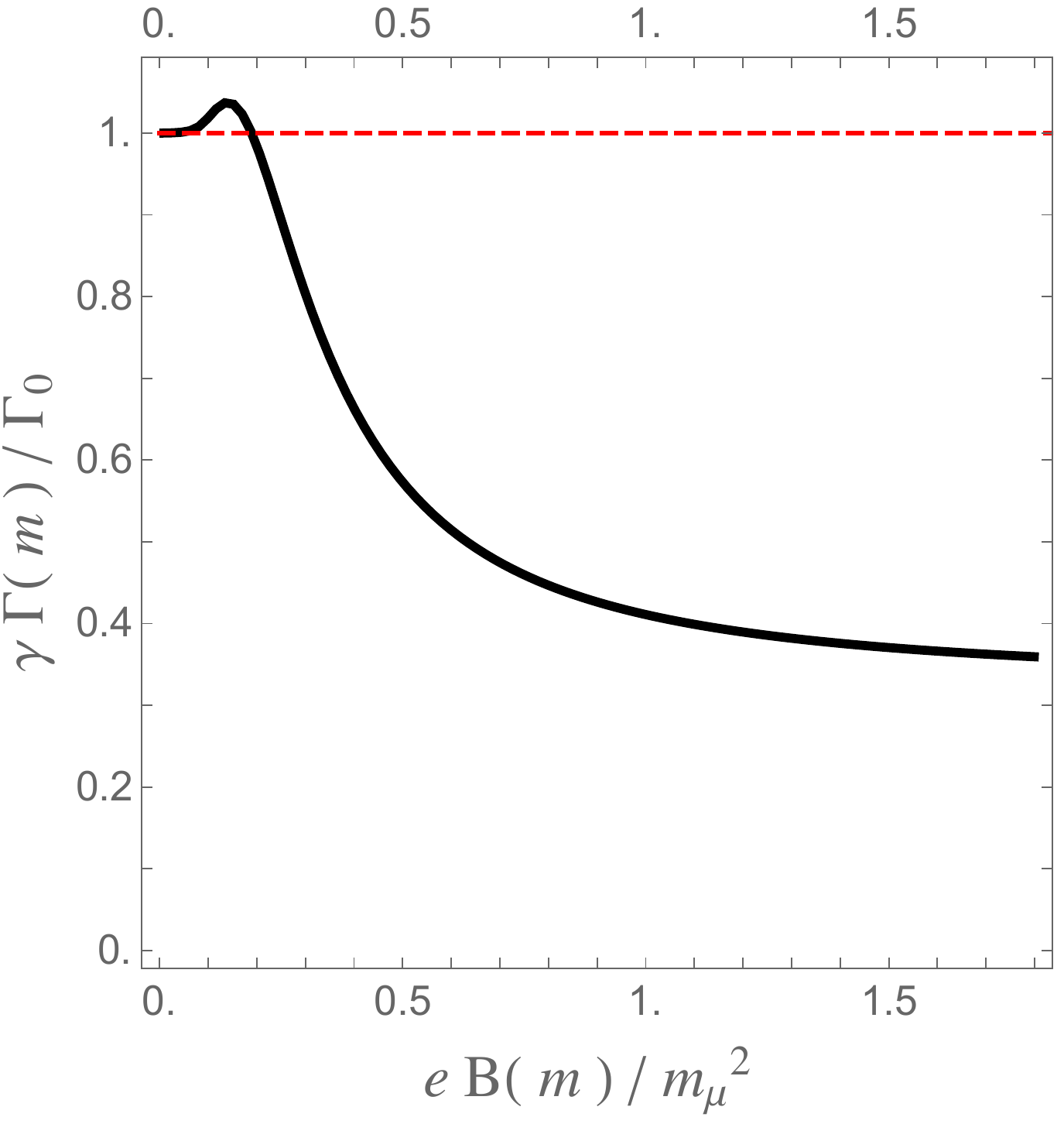}
\caption{Decay rate $\gamma\,\Gamma(m)/\Gamma'_0$ vs the magnetic field $B(m)$ for fixed radius $R=0.1\,\,\text{MeV}^{-1}=2\times 10^{-14}$ m.
Here $M_{\mu}=105.7$ MeV, $M_e=M_{\nu}=0$.}
\label{fig3}
\end{figure}

We consider the mass of the emitted particles to be much smaller than the mass of the decaying one, $M_e\approx M_{\nu}\approx 0$, and the initial momentum transverse to the magnetic field, $k^{\mu}_{z}=0$. 
In general, as can be seen from Eq. $(\ref{rad-clas})$ and Eq. $(\ref{cen-acc-c})$, when the radial energy $p_\perp$ of the incoming particle is fixed a larger magnetic field force it to move on a trajectory with a shorter radius and with a larger transverse acceleration. The quantum expression for the square of the radial energy is   
\b\label{radialq}
p_{\perp}^2=(2m+1)|e|B \,.
\e 
Now, assuming $p_{\perp}$ constant, the magnetic field can assume only discrete values given by the equation 
\b\label{B-vs-m}
|e|B=\frac{p_{\perp}^2}{2m+1}  \,,
\e
and therefore, higher values of $m$ are linked to larger radii and smaller accelerations. 
In Fig. $\ref{fig1}$ we show the behaviour of the ratio $\Gamma/\Gamma_0$ between the decay rates $(\ref{decayrate})$ and $(\ref{free-decay-rate})$ as a function of the Landau levels $m$ for different energies of the decaying particle, with the magnetic field given by the relation $(\ref{B-vs-m})$. The plot clearly shows that significant deviations from the inertial value start only at small Landau levels $m$, associated with larger accelerations. 
When the available energy is higher the particle's decay rate in the absence of magnetic field is approached later, as can be seen also in Fig. $\ref{fig2}$. As a matter of fact, for the same level $m$, a larger energy requires a stronger magnetic field to keep the particle along the same trajectory and the effect of the acceleration on the decay law is more important.
From the first of the expressions $(\ref{rad-vs-Llev})$, the magnetic field can be directly related to the Landau level and to the classical radius  
\b\label{eB}
|e|B=\frac{2m+1}{R^2} \,.
\e
Note that this equation can be obtained also equating the classical $(\ref{ene-class})$ with the quantum  $(\ref{ene-quant})$ expression for the energy. 
The Fig. \ref{fig3} shows the decay rate on the vertical axis and the magnetic field on the horizontal one. The magnetic field  is given by the expression $(\ref{eB})$ as a function of $m$ and with the radius kept constant.
 In this case, as the Landau level increases the magnetic field also increases and we can observe a major effect on the decay rate. The energy also changes according to the equation 
\b\label{ene-vs-Ll}
p_{\perp}=\frac{(2m+1)}{R} \,,
\e    
as can be seen inserting the Eq. $(\ref{eB})$ into $(\ref{radialq})$. 

\begin{figure}[ptb]
\includegraphics[width=3in]{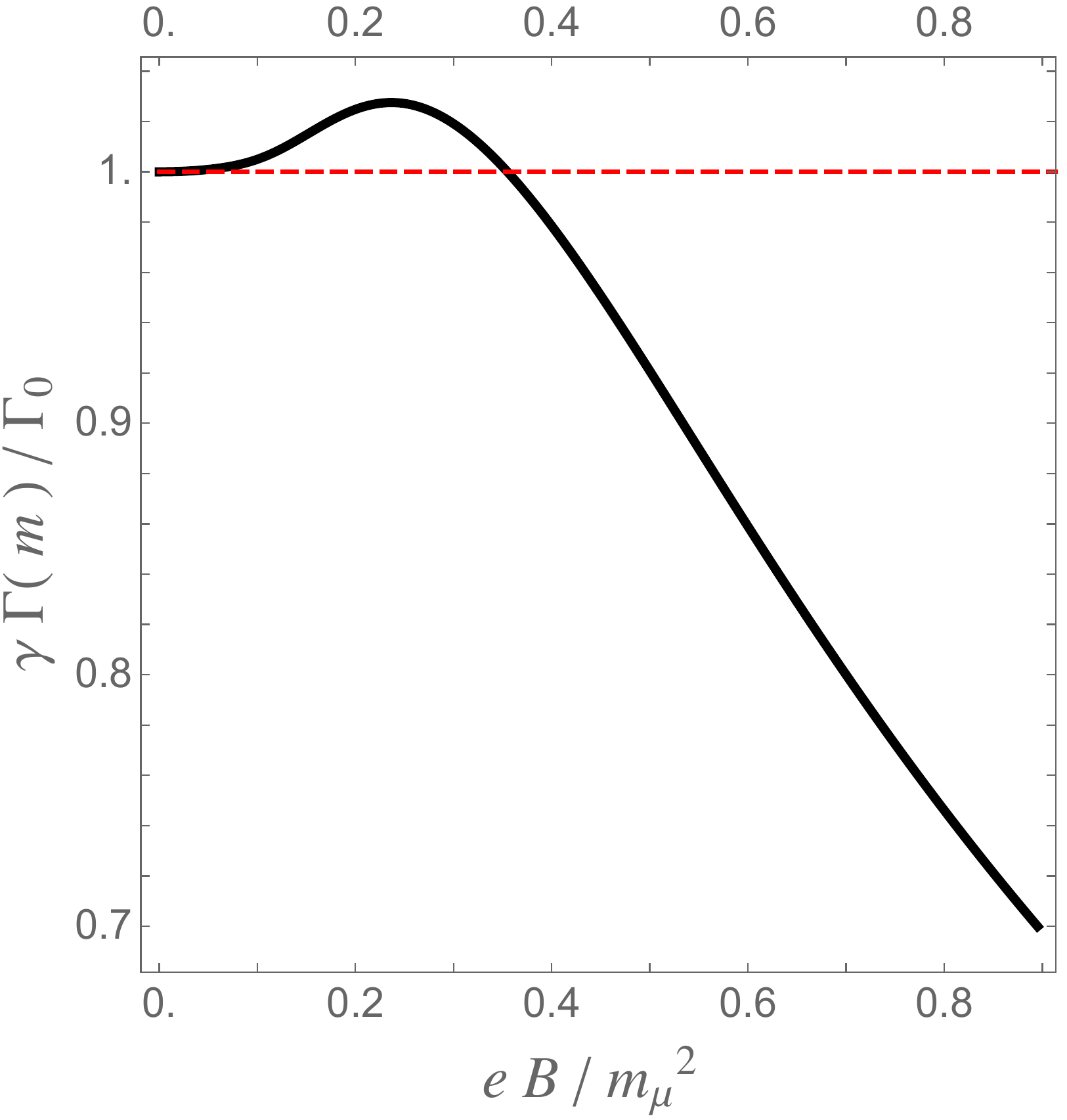}
\caption{Decay rate $\gamma\,\Gamma(m)/\Gamma'_0$ vs radial energy $p_\perp$ for fixed radius $R=0.1\,\, \text{MeV}^{-1}=2\times 10^{-14}$ m and $m=0$.
Here $M_{\mu}=105.7$ MeV, $M_e=M_{\nu}=0$.}
\label{fig4}
\end{figure}

It is not difficult to study the quantum limit for $m=n=0$ of the expression $(\ref{decayrate})$. The incoming particle occupies the zeroth Landau level when the radial energy squared equals the magnetic field, as can be easily seen from $(\ref{radialq})$. In this case the Laguerre polynomials are equal to one, $L_0^{\alpha}(x)=1$ \cite{GR}. The number of available Landau levels for the outgoing charged particle is limited by
\b
n\leq\frac{M_{\mu}^2}{2|e|B} \,.
\e   
For a magnetic field such that $|e|B>M_{\mu}^2/2$ the emitted particle would occupy the lowest Landau level $n=0$ and the decay rate ratio will reduce simply to 
\begin{align}
\frac{\gamma\,\Gamma}{\Gamma_0'}=&2\,\frac{e^{-(1+\frac{M_{\mu}^2}{2|e|B})}e^{-\sqrt{1+\frac{M_{\mu}^2}{|e|B}}}}{\sqrt{|e|B}}\nonumber\\
&\times\int_{0}^{x_{max}}dx\, \frac{e^{\sqrt{1+\frac{x^2}{|e|B}}}}{\sqrt{1+\frac{x^2}{|e|B}}} \,,
\end{align}
where $x=k_z^e$ and $x_{max}=\frac{M_{\mu}^2}{2\sqrt{M_{\mu}^2+|e|B}}$. From here, it can be seen that as $|e|B\to\infty$ the integration range shrinks and the decay rate goes to zero. 
In Fig. \ref{fig4} we show the rate for $m=0$ as a function of the magnetic field. Note that in this case the radial energy squared equals $|e|B$.

So far we have been using natural units, where $\hbar=1$ and $c=1$, now we want to associate specific values of the decay rate ratio to the relative radius and to the transverse acceleration in physical units, to make quantitative predictions.
We have the expressions
\begin{align}
R&=\frac{2m+1}{p_{\perp}}\times\hbar c  \,, \\
a_{\perp}&=\frac{p_{\perp}^{3}}{(2m+1)\omega_m^2}\times\frac{c}{\hbar} \,,
\end{align}  
where $\hbar$ and $c$ are the Planck constant and the speed of light.
Also, we might want to know the de Broglie wave-lenght $\lambda_{dB}$ associated to the particle, which is given by
\b
\lambda_{dB}=\frac{2\pi\,\hbar\,c}{p_{\perp}}   \,.
\e
In the following table, we show a few representative values of these observables 
\\
\begin{widetext}
\qquad\qquad\begin{tabular}{r|c|c|c|c|c|c|}\label{tab}
$\gamma\,\Gamma/\Gamma'_0 $ & $ p_{\perp}^2$ $[\text{MeV}^2]$   & Landau level $m$ & Radius [m] & Acceleration [m/$\text{s}^2$] & $\lambda_{dB}$ [m]  \\ \hline
$ 1.00094 $ &  $3\times 10^4$  &  $65$ & $1.49\times 10^{-13}$ & $ 4.39 \times 10^{29}$ &   $8.80 \times 10^{-15}$    \\ \hline
$ 1.0002 $ & $10^4$  &  $30$ & $1.20\times 10^{-13}$ &  $3.53 \times 10^{29}$ & $12.38 \times 10^{-15}$  \\ \hline   
$ 1.00008 $ & $5\times 10^3$ & $20$ & $1.14\times 10^{-13}$ &  $2.43 \times 10^{29}$ & $17.53 \times 10^{-15}$  \\ \hline                     
$ 1.00003 $ & $10^3$  &  $5$ &$ 6.86\times 10^{-14}$ &  $1.08 \times 10^{29}$ &$ 39.21 \times 10^{-15}$  \\ \hline                     
\end{tabular}
\end{widetext}
Note that a quantity of $|e|B\approx 10^{-2}\,M_{\mu}^2$ corresponds to a magnetic field $B\approx 10^{15}$ G. 

The lifetime of the muon can be measured with an accuracy of $10^{-5}$. For a detection of the induced modification of the decay rate, the effect has to be larger.  
In the paper \cite{Mue}, the authors studied the three body muon decay with scalar fields in the case of rectilinear acceleration and
it is interesting to observe that the acceleration predicted to have a potentially visible effect, $a = 7\times10^{27}g$, where $g\sim10\frac{m}{s^2}$ and $a$ is the proper acceleration of the decaying particle, is not far from the value we have found and reported in the last row of the table. We should note however, that the results are not  directly comparable because in their scenario only the unstable initial particle is accelerating, while in ours also the final charged particle is under acceleration.


\section{Concluding remarks}
The aim of this work was to investigate how realistic is the ideal clock hypothesis. We have considered the most fundamental possible clock, given by the lifetime of an unstable charged particle, and investigated how the decay rate is affected by a constant magnetic field, which imparts a centripetal acceleration to the charge. We have observed that significative deviations from the time dilation formula arise at huge accelerations, orders of magnitude further than the ones experienced by the particles in high energy physics experiments. As far as practical purposes are concerned, the ideal clock hypothesis is confirmed to be valid at the regimes where particle physicists normally operate or they plan to operate in the nearby future. On the other hand, our results confirm and extend the conclusions of \cite{LDL}, investigating a simpler and more straightforward scenario.

Some remarks regarding the limits of our model are in order. We have considered a toy model which does not describe the physical process of muon decay, where all the involved particles are fermions. This fact would imply a great complication due to the presence of the spin degrees of freedom. Furthermore, the interaction Lagrangian should be the weak force interaction which is not as elementary as the simple product of the fields we have considered. Still, those do not represent the main problem. The biggest difficulty with the real muon decay is that it involves three final particles, leading to three more integrals to be performed and more variables to deal with in the expression for the decay rate. We should note also that a strong magnetic field, as the one we need to have an observable effect in our model, can lead to pairs creation.

\acknowledgments
A. D. would like to thank Jason Doukas for valuable discussions at the early stage of this work. 
R.P and A.D. thank the National Science Centre, Sonata BIS Grant No. DEC-2012/07/E/ST2/01402 for the financial support. K.T. is partly supported by grant No. 2014/14/E/ST9/00152 from the National Science Centre.

\appendix


\begin{thebibliography}{99}
\bibitem{Bay}
J. Bailey et al., "Measurements of muon time dilation for positive and negative muons in a circular orbit", Nature 268, 301 (1977)
\bibitem{Har}
J. B. Hartle, "Gravity", Addison Wesley (2003)
\bibitem{Mue}
R. Mueller, "Decay of Accelerated Particles", Phys. Rev. D 56, 953-960 (1997)
\bibitem{VM}
D. A . T. Vanzella, G. E. A. Matsas, "Weak decay of uniformly accelerated protons and related process", Phys. Rev. D 63, 014010 (2000)
\bibitem{FVM}
D. Fregolente, G. E. A. Matsas, D. A . T. Vanzella, "Semiclassical approach to the decay of protons in circular motion under the influence of a gravitational field", Phys. Rev. D 74, 045032 (2006)
\bibitem{SY}
H. Suzuki, K. Yamada, "Analytic evaluation of the decay rate for an accelerated proton", Phys. Rev. D 67, 065002 (2003)
\bibitem{LDL}
K. Lorek, J. Louko, A. Dragan, "Ideal clocks - a convenient fiction", Class. Quantum Grav. 32, 175003 (2015)
\bibitem{FS}
D. H. Frisch, J. H. Smith, "Measurement of the Relativistic Time Dilation Using $\mu$-Mesons", Am. J. Phys. 31, 342 (1963)
\bibitem{GB}
V. G. Bagrov, D. M. Gitman, "Exact Solutions of Relativistic Wave Equations", Kluwer Academic Publisher (1990)
\bibitem{GR}
I. S. Gradshteyn, I. M. Ryzhik, "Table of Integrals, Series, and Products", Academic Press, New York (1980), formula $7.377$
\bibitem{Eis}
A. M. Eisele, "On the behaviour of an accelerated clock", Helv. Phys. Acta 60, 1024 (1987)


















\end{thebibliography}
\end{document}